\documentclass[11pt]{article}
\usepackage{amsmath,amssymb,color,graphics,epsfig,cite}

\usepackage{amsfonts}

\newcommand{\hoch}[1]{$\, ^{#1}$}

\makeatletter
\@addtoreset{equation}{section}
\makeatother

\newcommand{\be}{\begin{equation}}
\newcommand{\ee}{\end{equation}}
\newcommand{\bea}{\setlength\arraycolsep{2pt} \begin{eqnarray}}
\newcommand{\eea}{\end{eqnarray}}
\newcommand{\nn}{\nonumber}

\def\ft#1#2{{\textstyle{\frac{\scriptstyle #1}{\scriptstyle #2} } }}
\def\fft#1#2{{\frac{#1}{#2}}}

\def\0{{\sst{(0)}}}
\def\1{{\sst{(1)}}}
\def\2{{\sst{(2)}}}
\def\3{{\sst{(3)}}}
\def\4{{\sst{(4)}}}
\def\5{{\sst{(5)}}}
\def\6{{\sst{(6)}}}
\def\7{{\sst{(7)}}}
\def\8{{\sst{(8)}}}
\def\sst#1{{\scriptscriptstyle #1}}

\def\im{{{\rm i\,}}}
\def\R{{\mathbb R}}
\def\E{{\mathbb E}}

\begin{document}

\begin{flushright}
\hfill{MI-TH-1610}

\end{flushright}

\vspace{25pt}
\begin{center}
{\large {\bf Magnetically-Charged Black Branes and Viscosity/Entropy Ratios}}

\vspace{10pt}
Hai-Shan Liu\hoch{1,2}, H. L\"u\hoch{3} and C.N. Pope\hoch{2,4}

\vspace{10pt}

\hoch{1} {\it Institute for Advanced Physics \& Mathematics,\\
Zhejiang University of Technology, Hangzhou 310023, China}

\vspace{10pt}

\hoch{2} {\it George P. \& Cynthia Woods Mitchell  Institute
for Fundamental Physics and Astronomy,\\
Texas A\&M University, College Station, TX 77843, USA}

\vspace{10pt}

\hoch{3}{\it Department of Physics, Beijing Normal University,
Beijing 100875, China}

\vspace{10pt}

\hoch{4}{\it DAMTP, Centre for Mathematical Sciences,
 Cambridge University,\\  Wilberforce Road, Cambridge CB3 OWA, UK}

\vspace{40pt}

\underline{ABSTRACT}
\end{center}

We consider asymptotically-AdS $n$-dimensional black brane solutions in a theory of gravity coupled to a set of $N$ $p$-form field strengths, in which the
field strengths carry magnetic charges. For appropriately chosen
charges, the metrics are isotropic in the $(n-2)$ transverse directions.
However, in general the field strength configurations break the full
Euclidean symmetry of the $(n-2)$-dimensional transverse space. We
then study the linearised equation for transverse traceless metric
perturbations in these backgrounds, and by employing the Kubo formula we
obtain expressions for $\eta/S$, the ratio of shear viscosity to
entropy density.  We find that the KSS bound on the ratio $\eta/S$ is
generally violated in these solutions.  We also extend the discussion by
including also a dilatonic scalar field in the theory, leading to
solutions that are asymptotically Lifshitz with hyperscaling violation.

\vfill {\footnotesize Emails: hsliu.zju@gmail.com \ \ \ mrhonglu@gmail.com\ \ \
pope@physics.tamu.edu}

\thispagestyle{empty}

\pagebreak

\tableofcontents
\addtocontents{toc}{\protect\setcounter{tocdepth}{2}}



\section{Introduction}

   A class of static magnetically-charged planar black hole solutions of the
Einstein-Maxwell theory with a cosmological constant in an even spacetime
dimension $n$ was constructed in \cite{dhokkrau}.  In these solutions, the
Maxwell field strength takes the form $F=\alpha\,(dx_1\wedge dx_2 +
dx_3\wedge dx_4+\cdots)$, where $x_i$ are the coordinates on the
Euclidean $(n-2)$-dimensional transverse space of the planar black hole.
In the case of $n=4$ dimensions the solution is nothing but the standard
magnetically-charged Reissner-Nordstr\"om black hole with a planar
(or toroidal) horizon geometry.  In even dimensions $n\ge 6$ the solutions
are again asymptotic to AdS spacetime, but the magnetic charge
contribution in the $g_{tt}$ and $g^{rr}$ metric functions,
proportional to $1/r^2$, now falls off more slowly
than the black hole mass term, which falls like $1/r^{n-3}$.
This has significant implications for the properties of the boundary field theory related to the bulk theory via the AdS/CFT correspondence.

  In this paper, we shall generalise the discussion in \cite{dhokkrau}, by
considering an $n$-dimensional theory comprising a set of $N$ form
fields coupled to Einstein gravity with a (negative) cosmological
constant, possibly with the addition of a dilatonic scalar as well.  A
subclass of such theories that we shall initially focus
on comprises $N$ $p$-form field strengths coupled to gravity with a
cosmological constant.  We then study the asymptotically AdS
static planar black hole solutions that arise when the form fields carry magnetic charges.  A particular case within this class of theories and solutions is when $p=1$, meaning that the corresponding 1-form field strengths are simply the
gradients of a set of $N$ axionic scalar fields.  The black hole
solutions in this theory were constructed in \cite{andrade}, and
were studied recently in \cite{harramsan},
where properties such as the viscosity of the fluid described by the
boundary CFT were investigated in detail.

The AdS/CFT correspondence provides a new tool to study the dynamics of
strongly-coupled gauge field
theories \cite{adscft1,adscft2,adscft3,adscft4}, and
one of the most celebrated results is the universality, within a wide
class of theories, of the ratio of
shear viscosity to entropy density, namely
\be
\fft{\eta}{S} = \fft{1}{4 \pi} \,, \label{bound}
\ee
which was discovered, and further proposed as a lower bound, by
Kovtun, Son and Starinets (KSS) \cite{Policastro:2001yc,sonsta,KSS,KSS0}.
Recently, another viewpoint was highlighted, in which (\ref{bound})
is interpreted as a holographic boundary dual to a generalized
Smarr relation for the bulk gravitational backgrounds \cite{llpsma,liu}.

   It has been known for a while that the equality, and the KSS bound,
can be violated
under certain circumstances, notably in two different
situations.  The first of these is
if the theory involves higher-derivative gravity (see e.g.~\cite{kats,brlimyshya,caniohsu,cremonini},)
and the second is if the gravitational background has an anisotropic
transverse subspace (see e.g.~\cite{Natsuume:2010ky,Rebhan,Erdmenger:2010xm,mamo,Critelli:2014kra,
Bhattacharyya:2014wfa,Ge:2014aza}.\footnote{A violation of the viscosity bound has
also recently been observed in a Horndeski gravity theory, whose action
is purely linear in curvature \cite{fllp1, fllp2}.}) Recently,
it was observed in \cite{harramsan, alberte, burikham} that there is
a third way to violate the bound, in a  theory with conventional
two-derivative gravity whose gravitational
background is isotropic, but with $(n-2)$ scalar matter fields having
linear dependence on the $(n-2)$-dimensional
transverse coordinates.

   The solutions that we study in this paper provide
generalisations of those that were considered in
\cite{harramsan}, and we find that again the viscosities in these systems
can also violate the KSS bound.  There are significant differences
between the case of 1-form field strengths, as discussed in \cite{harramsan},
and the cases with higher-degree forms.  Principally, these differences
arise because now, unlike in the case of $p=1$, the isotropy of the
transverse space is broken by field strengths in the background solution,
and as a result different transverse traceless metric fluctuations can be
associated with boundary operators yielding different shear viscosities.
Some of the transverse and traceless (TT) modes give rise to $\eta/S$ ratios that violate the KSS bound, while others give the standard $1/(4\pi)$ result.

The paper is organized as follows.  In section 2, we consider AdS
black holes in $n=Np + 2$ dimensions carrying magnetic $p$-form charges.
In section 3, we study the linearised gravitational TT modes, and show
that they obey a Klein-Gordon equation with a position-dependent mass.
In section 4, we calculate the viscosity/entropy ratio, and evaluate it
in various limits and special cases.  We conclude the paper in section 5.
New classes of Lifshitz black holes with hyperscaling behaviour
in theories of gravity with form fields and a dilaton, where the
form fields carrying magnetic charges, are given in an appendix.

\section{AdS planar black holes with magnetic charges}\label{adsbhsec}

In this section we consider Einstein gravity coupled to $N$ $p$-form field
strengths $F^I_{(p)}=dA^I_{(p-1)}$, $I=1,2,\ldots, N$, with the Lagrangian
\be
{\cal L} = \sqrt{-g} \Big(R -2\Lambda -
\fft{1}{2\, p!} \sum_{I=1}^N (F_{(p)}^I)^2\Big)\,.\label{Npformlag}
\ee
A general class of AdS black branes were constructed in \cite{Bardoux:2012aw}.  For simplicity, we assume that the spacetime dimension is
\be
n=N p + 2\,.
\ee
The theory admits AdS planar black hole solutions with the metric ansatz
given by
\be
ds_n^2 = -h\, dt^2 + \fft{dr^2}{f} + r^2 d\Sigma^2\,,\label{metricans}
\ee
where $d\Sigma^2$ is the metric on an $(n-2)$-dimensional Euclidean space.
The Einstein tensor $G_{\mu\nu}=R_{\mu\nu}-\ft12 R g_{\mu\nu}$ has non-zero
components given by
\bea
G_{tt} &=& -\fft{(n-2)}{2r}\, h\, f' - \fft{(n-2)(n-3)}{2 r^2}\, h f\,,\nn\\
G_{rr} &=& \fft{(n-2)}{2r}\, \fft{h'}{h} + \fft{(n-2)(n-3)}{2r^2}\,,\\
G_{ij} &=& \Big[\fft{(n-3)r}{2}\, \Big(f' +\fft{f\, h'}{h}\Big) +
\fft{r^2\, f h''}{2h} +\fft{r^2\, f' h'}{4h} -\fft{r^2 \, f {h'}^2}{4 h^2}
  +\ft12 (n-3)(n-4) \, f\Big]\, \delta_{ij}\,.\nn
\eea

    Here, we shall consider the case where the $N$ $p$-form field strengths
each carry a magnetic charge and each spans a disjoint $p$-dimensional
subspace of the $(Np)$-dimensional Euclidean space.
Since the metric ansatz (\ref{metricans}) is isotropic in the $(n-2)$
transverse directions, the $N$
magnetic charges must necessarily be equal. Thus we may write
\bea
d\Sigma^2 &=& d\Sigma_1^2 + d\Sigma_2^2 + \cdots d\Sigma_N^2\,,\cr
d\Sigma_1^2 &=& dx_1^2 + \cdots + dx_p^2\,,\qquad
d\Sigma_2^2 = dy_1^2 + \cdots + dy_p^2\,,\qquad \hbox{\it etc.}\,,
\eea
with the magnetically charged field strengths given by
\bea
F_{(p)}^I &=& \alpha \Sigma_{(p)}^I\,,\qquad I=1,2,\ldots,N\,,\cr
\Sigma_{(p)}^1 &=& dx_1\wedge\cdots \wedge dx_p\,,\qquad
\Sigma_{(p)}^2 = dy_1\wedge \cdots \wedge dy_p\,,\qquad \hbox{\it etc.}
\label{field}
\eea
The black hole solution is then given by
\be
h=f = g^2 r^2 -\fft{\alpha^2}{\Delta\,r^{2p-2}} -
\fft{\mu}{r^{n-3}}\,,\qquad \Delta=2p(n-2p -1)=2(N-2)p^2+2p\,,\label{hfsol}
\ee
where $\ell=1/g$ is the AdS radius, defined by
\be
g^2 =-\fft{2\Lambda}{(n-1)(n-2)}\,.
\ee
Note that the $p=1$ case corresponds to the linear axion models \cite{andrade}.

  The $(n-2)$-dimensional metric $d\Sigma^2$ on the transverse
space has the full isometries of the $(n-2)$-dimensional Euclidean
group $\E^{n-2}$,
namely, the semi-direct product of the translations $\R^{n-2}$ with the
rotation group $SO(n-2) =SO(Np)$.  The ansatz (\ref{field})
for the $N$ $p$-form field strengths, on the other hand, breaks the
$SO(Np)$ rotational symmetry of the complete solution down to $SO(p)^N$.
Since the gauge potentials $A^I_{(p-1)}$ enter the equations of motion
only via their gauge-invariant field strengths $F^I_{(p)}=dA^I_{(p-1)}$,
we may view the translational symmetries of the $(n-2)$-dimensional
transverse space as remaining unbroken in the bulk solution, modulo gauge
transformations.\footnote{One could take, for example, $A^1_{(p-1)}=
 \alpha\, x_1\, dx_2\wedge \cdots \wedge dx_p$, which is
not itself translationally invariant under $x_1\longrightarrow x_1 + c_1$,
but it is invariant modulo the gauge transformation
$A^1_{(p-1)}\longrightarrow A^1_{(p-1)} + d\Lambda^1_{(p-2)}$, with
$\Lambda^1_{(p-2)}= -c_1\, x_2\, dx_3\wedge \cdots \wedge dx_p$.  Note
that although the bulk solutions can be viewed as being translationally
invariant, translational symmetry is broken in the boundary CFTs, since the
potentials themselves are the sources dual to operators in the boundary
theory. A review of theories with broken translation invariance can be
found in \cite{taywoo}.}
The case when $p=1$ is special; this corresponds to $N$ 1-form
field strengths $F^I_{(1)}= d\chi^I$.  The scalar Lagrangian has a global
$\E^N$ symmetry, comprising the semi-direct product of the axionic
shift symmetries $\R^N$ and the internal $SO(N)$ rotations.  Setting
$\chi^I = \alpha x^I$ equates the $\E^N$ internal scalar symmetries with the
$\E^N$ symmetries of the Euclidean transverse
metric $dx^I dx^I$, leading to a spontaneous
breaking of $\E^N\times \E^N$ to the diagonal $\E^N$ subgroup.  As far as
the bulk solution is concerned, one could choose to attribute this to a
breaking of the internal scalar symmetries, with the Euclidean
symmetries of the transverse space remaining unbroken.  However,
from the point of view of the
CFT, the explicit coordinate dependence of the scalars
$\chi^I$ implies that the dual operators in the boundary theory break
the translational symmetries.

For $p\ge 2$, the metric (\ref{metricans}) with (\ref{hfsol}) can instead be supported by a lesser number of field strengths.  For example, we can use just
one $p$-form field strength, with
\be
F_{(p)}^1=\alpha \big(\Sigma^1_{(p)} + \Sigma^2_{(p)} + \cdots +
\Sigma_{(p)}^N\big)\,.\label{field2}
\ee
The $p=2$ solution with a single Maxwell field was constructed in 
\cite{dhokkrau}.
The general $p$-form solutions were previously constructed in 
\cite{Bardoux:2012aw}. (See also \cite{Ortaggio:2007hs,Ortaggio:2014gma}.)  
As we shall see later, although the black hole metric is the same for
these different form-field configurations, a transverse traceless
metric perturbation can, under certain circumstances, be different, leading
to a different viscosity/entropy ratio.  For now, we note that
for $p\ge3$ the expression (\ref{field2}) is invariant under
the $SO(p)^N$ subgroup of the $SO(Np)$ rotational symmetry group
of the transverse Euclidean space.  The case $p=2$ is special,
as we shall discuss later, with the symmetry now becoming $U(N)$.

It is worth pointing out that in our construction or in that of 
\cite{dhokkrau}, the spacetime dimensions are forced to be $D=N p + 2$ for 
given $p$-forms.  In particular this rules out $D=5$ for $p=2$.  However, 
isotropic AdS black branes with magnetic charges can be constructed 
\cite{Donos:2011qt} in the $U(1)^3$ gauged supergravity that can be obtained 
from the $S^5$ reduction of type IIB supergravity \cite{Cvetic:1999xp}.

\section{Transverse, traceless perturbations}

   We now consider a transverse, traceless, perturbation of the
background, by replacing the metric (\ref{metricans}) by\footnote{For now,
we consider a perturbation $\delta g_{xy}$ involving just a single
component of the metric.  Later, we shall encounter a case where it
is necessary to consider perturbations with two non-zero components,
$\delta g_{x_1 y_1}$ and $\delta g_{x_2 y_2}$.}
\be
ds^2 = -h\, dt^2 + \fft{dr^2}{f} +
   r^2 (d\Sigma_{n-2}^2 + 2 \Psi(r,t)\, dx dy)\,.\label{pertmet}
\ee
Here, $x$ and $y$ represent, for now, any two out of the total set of
$(n-2)$ coordinates in the Euclidean space of the $d\Sigma_{n-2}^2$ metric.
The original Ricci tensor of the background metric acquires a single
non-vanishing correction at linear order in $\Psi$, namely
\bea
R^{(1)}_{xy} &=& -\ft12 r^2 \, \square\Psi - \ft12 r f'\, \Psi -
   \fft{r f h'}{2h}\, \Psi -(n-2) f\, \Psi\,,\nn\\
&=&  -\ft12 r^2 \, \square\Psi + R^{(0)}_{xx}\, \Psi\,,
\eea
where $R^{(0)}_{xx}$ denotes any of the (diagonal) components of the
zeroth-order Ricci tensor of the background metric, in the directions
of the  transverse $(n-2)$-dimensional Euclidean space.  Thus the
Einstein equation $G_{\mu\nu}=\ft12 T_{\mu\nu}$, which can be graded in
orders of powers of $\Psi$ to give $G^{(0)}_{\mu\nu}=\ft12 T^{(0)}_{\mu\nu}$,
$G^{(1)}_{\mu\nu}=\ft12  T^{(1)}_{\mu\nu}$ and so on, leads to
\bea
G^{(1)}_{xy} &=& -\ft12 r^2 \, \square\Psi +
        R^{(0)}_{xx}\, \Psi -\ft12 R^{(0)}\, g^{(1)}_{xy}\,,\nn\\
&=& -\ft12 r^2 \, \square\Psi + G^{(0)}_{xx}\, \Psi\,,\nn\\
&=&  -\ft12 r^2 \, \square\Psi + \ft12 T^{(0)}_{xx}\, \Psi\,,
\eea
and hence
\be
 \square\Psi - \fft{1}{r^2}\, (T_{xx}^{(0)}\, \Psi - T^{(1)}_{xy})=0\,.
\label{boxPsi}
\ee
Writing $T_{xy}^{(1)} = (\delta T_{xy}/\delta g_{xy})\, \delta g_{xy}$ gives,
as in \cite{harramsan},
\be
\square\Psi -m^2(r)\, \Psi=0\,,\label{TTeqn}
\ee
where
\be
m(r)^2 = g^{xx}\, T_{xx} - \fft{\delta T_{xy}}{\delta g_{xy}}\,,
\label{msquared}
\ee
which reproduces the result in \cite{harramsan}.
In the metric background (\ref{metricans}), eqn (\ref{TTeqn}) is given by
\be
f \Psi'' + \Big(\fft{f h'}{2h} +\ft12 f' + \fft{(n-2)}{r}\, f\Big) \Psi'
-\fft1{h}\, \ddot\Psi- m(r)^2 \Psi =0 \,,\label{TTeqn2}
\ee
where a dot denotes a derivative with respect to $t$.

   The energy-momentum tensor for the set of $N$ $p$-form field strengths
is given by
\be
T_{\mu\nu} =\fft1{(p-1)!}\, \sum_{I=1}^N \Big[ F^I_{\mu\rho_2\cdots \rho_p}
\, F^I_\nu{}^{\rho_2\cdots \rho_p} - \fft1{2p}\, (F^I)^2\, g_{\mu\nu}\Big]\,.
\label{Tmunu}
\ee
With the fields given by (\ref{field}), we therefore have that in the
background solution,
\be
T_{x_1 x_1} = T_{x_2 x_2}=\cdots = T_{y_1 y_1} = T_{y_2 y_2}=\cdots
  = \fft{\alpha^2\, (1-\ft12 N)}{r^{2p-2}}\,.\label{Tdiag}
\ee
If we consider a TT metric fluctuation such as $\delta g_{x_1 y_1}$ (or any
other case where one index is chosen from the range of coordinates
spanned by one of the field strengths, and the other index chosen from the
range of coordinates spanned by another field strength), then the
first-order fluctuation in $T_{\mu\nu}$ will come purely from the fluctuation
of $g_{\mu\nu}$ in the $F^2 g_{\mu\nu}$ terms in (\ref{Tmunu}), and hence
we shall have
\be
T_{x_1 y_1} = -\fft{\alpha^2\, N}{2 r^{2p}}\, \delta g_{x_1 y_1}\,.
\ee
Thus, substituting into the expression (\ref{msquared}), we find
\be
m(r)^2 = \fft{\alpha^2}{r^{2p}}\,\label{msqxy}
\ee
This is the generalisation, to the case of $N$ $p$-form field strengths, of
the $N$ axionic scalars considered in \cite{harramsan}, which can be viewed
as the potentials for 1-form field strengths.

   An alternative possibility for the TT perturbation, keeping the
same configuration (\ref{field}) for the field strengths, is to take both
of the indices on the metric fluctuation to lie in a range spanned by
a single field strength.  (This requires that $p\ge 2$.) Let us consider, 
for example, $\delta g_{x_1 x_2}$.  The first-order
fluctuation of the energy-momentum tensor will now receive contributions
from both the terms inside the summation in (\ref{Tmunu}), giving
\be
T_{x_1 x_2}= \fft{\alpha^2\, (1-\ft12 N)}{r^{2p}}\, \delta g_{x_1 x_2}\,.
\ee
Substituting this result, together with (\ref{Tdiag}), into (\ref{msquared})
then gives
\be
m(r)^2 =0\label{msqxx}
\ee
for this case.

  One can also consider other possible form-field configurations that
will give rise to black hole solutions with the metric taking the
spatially-isotropic form (\ref{metricans}), such as the case of a
single $p$-form field strength spanning an $NP$-dimensional
transverse space, as in (\ref{field2}).  It is easy to verify that if
$p\ge 3$, this gives the same results for $m(r)^2$ as we obtained in the
previous example with $N$ distinct $p$-forms each spanning non-overlapping
$p$-dimensional subspaces.  Namely, with the ansatz (\ref{field2}) for
a single $p$-form, we find that $m(r)^2$ is given by (\ref{msqxy}) if the
two indices on the metric perturbation are taken from two different
$p$-dimensional subspaces, and that $m(r)^2$ vanishes as in
(\ref{msqxx}) if the two indices lie within the same $p$-dimensional subspace.

   A special case arises if a 2-form field strength spans a $2N$-dimensional
transverse space, as in (\ref{field2}). Now, unlike the situation when $p\ge3$,
the term $F_{\mu\rho_2\cdots \rho_p} \, F_\nu{}^{\rho_2\cdots \rho_p}$
in the energy-momentum tensor will also give a contribution at linear order
in a metric perturbation $\delta g_{xy}$ where the two indices lie
in different 2-dimensional subspaces.  Thus if the transverse
directions are $(x_1,x_2,y_1,y_2,\cdots )$, and we consider a metric
fluctuation $\delta g_{x_1 y_1}$, then there will be linear-order
terms in the energy momentum tensor given by
\be
T^{(1)}_{x_1 y_1}= -\fft{N\, \alpha^2}{2 r^4}\, \delta g_{x_1 y_1}\,,\qquad
T^{(1)}_{x_2 y_2} = -\fft{\alpha^2}{r^4}\, \delta g_{x_1 y_1}\,.
\ee
In order to obtain TT eigenfunctions of the linearised fluctuation
equations, it is therefore necessary to consider metric fluctuations
with both $\delta g_{x_1 y_1}$ and $\delta g_{x_2 y_2}$ non-zero in this case.
The corresponding TT Lichnerowicz modes are diagonalised by writing
$\delta g_{x_1 y_1}= \ft12 r^2 (\Psi_+  + \Psi_-)$ and $\delta g_{x_2 y_2}=
\ft12 r^2 (\Psi_+ - \Psi_-)$, for which, using also the appropriate extension
of (\ref{boxPsi}), we find
\be
\square\Psi_\pm - m_\pm(r)^2\, \Psi_\pm=0\,,
\ee
with
\be
m_+(r)^2 = \fft{2\alpha^2}{r^4}\,,\qquad m_-(r)^2 =0\,.
\label{msqpm}
\ee

   It is worth remarking that the single 2-form field with the
ansatz (\ref{field2}) in the $2N$-dimensional transverse space also
breaks the $SO(2N)$ rotational symmetry of the background metric.  In
this case, it is broken down to $U(N)$, as may be seen by introducing
complex coordinates $z_a$ defined by
\be
z_1 = x_1 + \im\, x_2\,,\qquad z_2 = y_1 + \im\, y_2\,,\qquad \hbox{etc.}
\,,
\ee
in terms of which the field strength can  be written as
\be
F_{(2)}= \fft{\im\, \alpha}{2}\, dz_a\wedge d\bar z_a\,.
\ee
Thus $F_{(2)}$ can be seen to be $\alpha$ times the K\"ahler form on
the $2N$-dimensional Euclidean space, and it manifestly has a
$U(N)$ symmetry.

\section{Viscosity/entropy ratio}

In the previous sections, we consdered magnetically charged black branes 
and analyzed the transverse and traceless perturbation modes. Now, we turn 
to the derivation of the viscosity to entropy density ratio. Usually, it is 
a universal value (\ref{bound}) for gauge theories which have  
two-derivative Einstein gravity dual. However, as is observed  
in \cite{harramsan},
the situation is changed when the theory involves anisotropic matter fields.

\subsection{A general discussion}

As discussed in \cite{harramsan}, and the previous section,
a traceless and transverse mode in
the metric (\ref{metricans}) satisfies (\ref{TTeqn}).
Consider $\Psi(r,t)=e^{-{\rm i}\omega t} \psi(r)$.  Near the horizon
with
\be
h =h_1\, (r-r_0) + h_2\, (r-r_0)^2 +\cdots\,,\qquad
f = f_1\, (r-r_0)+ f_2\, (r-r_0)^2+\cdots\,,
\ee
we have
\be
\psi(r) \sim  \exp\Big(-\fft{{\rm i}\omega}{4\pi T}\, \log(r-r_0)\Big)\,,
\ee
where
\be
T=\fft{\sqrt{f_1 h_1}}{4\pi}
\ee
is the Hawking temperature.
This leads to the ansatz for $\psi$ in a small-$\omega$ expansion, with
\be
\psi(r) = \chi(r)\, \exp\Big(-\fft{{\rm i}\omega}{4\pi T} \log f(r)\Big)
(1 - {\rm i\omega}\,U(r)) + {\cal O}(\omega^2)\,,
\ee
where $\chi$ is the solution of equation (\ref{TTeqn2}) with $\omega=0$:
\be
f \chi'' + \Big(\fft{f h'}{2h} +\ft12 f' + \fft{(n-2)}{r}\, f\Big) \chi'
- m(r)^2 \chi =0\,.\label{chieqn}
\ee
For the massless case with $m(r)=0$, $\chi$ is simply a constant, and can
be set to be 1.  For non-vanishing $m(r)$, $\chi$ depends on $r$.
If $m(r)$ falls off sufficiently fast asymptotically,
then $\chi\rightarrow 1$ as $r\rightarrow \infty$.  The function $U$ can also
be solved up to quadratures, and is given by
\be
U'=\fft{1}{r^{n-2} \chi^2 \sqrt{h f}}
\Big(c - \fft{\chi^2 f' r^{n-2}}{4\pi T}\,\sqrt{\fft{h}{f}}\Big)\,,
\ee
where the integration constant $c$ should be chosen such that
the zero in the denominator coming from the vanishing of $f$ and $h$ on
the horizon is cancelled. Thus we have
\be
c=r_0^{n-2} \chi(r_0)^2=4 S\, \chi(r_0)^2\,.
\ee
where $S$ is the entropy density, given by
\be
S=\ft14 r_0^{n-2}\,.
\ee
This implies
\bea
\psi^* \psi' =  - \fft{4{\rm i}\omega S \chi(r_0)^2}{\sqrt{hf}\, r^{n-2}}
+\chi\chi'  + {\cal O}(\omega^2)\,.
\eea
Following the procedure described in \cite{sonsta,brlimyshya}, and
applied to black holes with the metric ansatz (\ref{metricans}) in
\cite{llpsma}, the
shear viscosity will then be given by
\be
\eta = -\lim_{r\to\infty}
    \Big[\fft{r^{n-2} \sqrt{hf}\, (\psi^*\psi'-\hbox{c.c.})}{32\, \im
  \pi\omega}\Big]
= \fft{S}{4\pi}\, \chi(r_0)^2\,,
\ee
and hence
\be
\fft{\eta}{S} = \fft{\chi(r_0)^2}{4\pi}\,.\label{etaoverS}
\ee

We now apply the above discussion to the AdS planar black holes in
section \ref{adsbhsec}.  The metric functions were given in
(\ref{hfsol}), and $m(r)$ is given by
\be
m(r)^2 = \fft{\nu\,\alpha^2}{r^{2p}}\,.\label{generalmass}
\ee
The parameter $\nu$ can be 0, 1 or 2, corresponding to the cases we saw in
(\ref{msqxx}), (\ref{msqxy}) and the $m_+(r)^2$ expression in
(\ref{msqpm}) respectively. (The $\nu=2$ case arises only when $p=2$.)

\subsection{The special case $\mu=0$}

   We first consider the case where $\mu=0$ in the metric functions
given in (\ref{hfsol}).  The horizon is located at $r_0$ with
\be
r_0^{2p} = \fft{\alpha^2}{\Delta g^2}\,,
\ee
and the Hawking temperature is
\be
T=\fft{p g^2}{2\pi \Delta^{\fft1{2p}}} \big(\fft{\alpha}{g} \big)^{\ft{1}{p}}
\,.
\ee
The function $\chi$ can be found explicitly in this case, and is given by
\bea
\chi &= &  c_1 \, {}_2F_1[\ft12(1-x-y), \ft12(1-x + y); 1-x; \big(\ft{r_0}{r}\big)^{2p}] \cr
&&+
\fft{c_2}{r^{2px}}\, {}_2F_1[\ft12(1+x-y), \ft12(1+x + y); 1+ x; \big(\ft{r_0}{r}\big)^{2p}]\,,
\eea
where $c_1$ and $c_2$ are two integration constants and
\be
x=\fft{n-1}{2p}\,,\qquad y=\fft{\sqrt{(n-2p-1)(n-2(4\nu+1) p-1)}}{2p}\,.
\ee
The quantity $\chi$ diverges when $r=r_0$ for general values of
$c_1$ and $c_2$, but it can be made convergent by choosing
\be
\fft{c_2}{c_1} = -\fft{\Gamma(1-x)\Gamma(\fft12(1+x-y))
  \Gamma(\fft12(1+x+y))}{
\Gamma(1+x)\Gamma(\fft12(1-x-y))\Gamma(\fft12(1-x + y))} \,
\fft{1}{r_0^{2px}}\,.
\ee
With this choice, we find
\be
\chi(r_0) = \fft{2c_1\pi \Gamma(1-x)\sin(\pi x)}{(\cos(\pi x) +
       \cos(\pi y))\Gamma(\ft12(1-x-y))\Gamma(\ft12(1-x +y)}\,.
\ee
Thus for this case, we have
\be
\fft{4\pi\eta}{S} =  \Big(\fft{2\pi \Gamma(1-x)\sin(\pi x)}{(\cos(\pi x)
+ \cos(\pi y))\Gamma(\ft12(1-x -y))\Gamma(\ft12(1-x + y))}\Big)^2\,,
\ee
which is less than $1$.  The result with $p=1$ in four dimensions
was obtained previously in \cite{Davison:2014lua}.

   Turning now to the case where the parameter $\mu$ in the metric functions
$h$ and $f$ in (\ref{hfsol}) is non-zero, it is no longer possible to obtain a
general closed-form expression for $\chi(r)$. It is nevertheless possible to
obtain expansions valid for small values of the magnetic charge
parameter $\alpha$, or else for small values of the Hawking temperature
of the black hole, as we shall discuss below.

\subsection{Small $\alpha$ (large $T$) expansion}

   The metric functions (\ref{hfsol}), and the other quantities needed
for solving (\ref{chieqn}), are given by
\be
h=f=g^2 r^2 - \fft{\alpha^2}{\Delta\,r^{2p-2}} - \fft{\mu}{r^{n-3}}\,,\qquad
m(r)^2 = \fft{\nu\, \alpha^2}{r^{2p}}\,,\qquad \Delta = 2p(n -2p - 1)\,.
\ee
The horizon is located at $r=r_0$, implying that $\mu$ is given by
\be
\mu=g^2 r_0^{n-1}\Big(1 - \fft{\alpha^2}{\Delta g^2 r_0^{2p}}\Big)\,,
\ee
and hence the Hawking temperature $T=h'(r_0)/(4\pi)$ is given by
\be
T=  \fft{(n-1) g^2\, r_0}{4\pi} \,
\Big[1 -\fft{\alpha^2}{2p(n-1)g^2\, r_0^{2p}} \Big] \,.
\ee
Defining $T_0$ to be the temperature when $\alpha=0$, we have
\be
T=T_0\Big[ 1 -\fft1{8\pi p}\,
\Big(\fft{(n-1) g^2}{4\pi}\Big)^{2p-1}\, \Big(\fft{\alpha}{T_0^p}\Big)^2\Big]
\,,\qquad T_0= \fft{(n-1) g^2\, r_0}{4\pi}\,.
\ee
If $\alpha/T_0^p<<1$ then we have $T\approx T_0$, with $T\le T_0$, and so
if $\alpha/T^p<<1$ we also have $\alpha/T_0^p<<1$.

    We may seek a solution for the function $\chi$ as a series expansion
in powers of $\alpha$, writing
\be
\chi = 1 + \alpha^2 \tilde \chi + \cdots\,.
\ee
Substituting into (\ref{chieqn}), we find that $\tilde\chi$ has the solution
\be
\tilde \chi(z) = -\fft{\nu}{g^2(n-1-2p)r_0^{2p}}\Big[\Big(\fft{z^{2p}}{2p}\Big)
\, {}_2F_1[1, \ft{2p}{n-1}; 1 + \ft{2p}{n-1}; z^{n-1}] +
\fft{1}{n-1} \, \log(1 - z ^{n-1})\Big] \,,
\ee
where $z=r_0/r$. Thus on the horizon, at $z=1$, we have
\be
\tilde \chi(1) =
\fft{\nu\, [\gamma + P(\ft{2p}{n-1})]}{(n-1)(n-2p-1) g^2 r_0^{2p}}\,,
\ee
where $\gamma$ is the Euler-Mascheroni constant and
$P(z)\equiv \Gamma'(z)/\Gamma(z)$ is the digamma function.  This
implies that the viscosity/entropy ratio is given by
\be
\fft{4\pi \eta}{S} = 1 + \fft{2\nu (n-1)^{2p-1}\,
    g^{4p-2}}{(n-2p-1)(4\pi)^{2p} }\, \big[\gamma +
P(\ft{2p}{n-1})\big]\, \Big(\fft{\alpha}{T^p}\Big)^2  +
{\cal O}\Big(\big( \fft{\alpha}{T^p}\big)^4 \Big)\,.
\ee
Note that the value of $\gamma + P(\ft{2p}{n-1})$ is
always less than zero (for the relevant values of $n$ and $p$),
so the ratio $4\pi \eta/S$ is less than $1$.  The result for
the special $n=4$ and $p=1$ case was obtained in \cite{harramsan}.

\subsection{Low temperature expansion}

When the magnetic charge parameter $\alpha$ is given by
\be
\alpha^2 = 2(n-1)p g^2 r_0^{2p}\,,
\ee
the solution becomes extremal with zero temperature.  In this case, viscosity
vanishes.  At the low-temperature, we find that the equation for $\chi$
seemingly cannot be solved analytically, except when $p=1$, with $n=4$ or 5.
The four-dimensional result was given in \cite{harramsan}. In five
dimensions (with $\nu=1$), we find
\be
\chi(z)=\fft{1-3z^2}{1-z^2} - \fft{4z^4\log z}{(1-z^2)^2} +
\fft{2\pi T z^2(1-z^4 +4z^2 \log z)}{g^2 r_0 (1-z^2)^3} + {\cal O}(T^2)\,,
\ee
where again, $z=r_0/r$.
This means that at on the horizon (i.e. at $z=1$),
\be
\chi(1)= -\fft{2\pi T}{3 g^2 r_0}\,,
\ee
and hence, from (\ref{etaoverS}), we find
\be
\fft{4\pi \eta}{S} = \fft{4\pi^2 T^2}{9g^4 r_0^2} + \cdots =\fft{32 \pi^2}{9g^2} \big(\fft{T}{\alpha}\big)^2 + {\cal O}\Big(\big(\fft{T}{\alpha}\big)^4\Big)\,.
\ee
Thus the viscosity bound (\ref{bound}) is violated in the
special case when $\mu = 0$, and in the high-temperature limit, and also
in the low-temperature limit where
the viscosity/entropy ratio is proportional to $T^2$.  The low-temperature
result provides further examples of the phenomenon
seen already in \cite{harramsan} and elsewhere, that the ratio $\eta/S$ can
not only be less than the KSS bound $1/(4\pi)$, but it can be arbitrarily
close to zero.\footnote{A large violation of the KSS bound in a system
without pathologies was first exhibited in \cite{buccre}.}

\section{Conclusions}

   In this paper, we studied $n$-dimensional
magnetically-charged planar black hole
solutions in the theory of Einstein gravity coupled to $N$  $p$-form field
strengths. In general,  turning on the magnetic charges would result in
solutions where the metric in the transverse space would be
anisotropic.  However, by  requiring the field strengths to span the
entire transverse space, with equal magnetic charges associated with each
of the $p$-dimensional planes, we obtained solutions where the
metric was isotropic in the transverse dimensions, with the full Euclidean
$\E^{n-2}$ symmetry.  The field strengths in the solution in
general break the associated
$SO(n-2)$ rotational symmetry of the metric to a subgroup.

    We then studied the transverse and traceless metric perturbations of
the background solutions.  Although the metric is isotropic,
the linearised TT modes turn out to be satisfy scalar wave equations with
different radially-dependent ``mass terms'' that are proportional to the
magnetic charge parameter $\alpha$ as in (\ref{generalmass}). For $p=1$
forms, corresponding to axion potentials, we have $\nu=1$; for $p=2$
Maxwell field strengths, we have $\nu=0$, 1 or 2, depending on the
nature of the charge configurations; for $p\ge 3$, we have $\nu=0$ or 1.
The viscosity can
be calculated by means of the Kubo formula, and it turns out to depend on the
magnitude of the zero-frequency linearised mode evaluated on the
horizon. It can be easily established that the $m(r)=0$ cases yield the
standard KSS ratio $\eta/S=1/(4\pi)$.  For $m(r)^2> 0$, the general
linearised equation cannot be solved analytically, although there is a
special case where the mass parameter $\mu$ in the black hole solution
vanishes, for which we can obtain an
analytical solution for the linearised mode.  We then studied the cases
where $\mu\ne0$, for which we could solve the linearised equation in
the limits of a high-temperature expansion and also in a low-temperature
expansion. In both cases, the viscosity bound is violated, and
the ratio $\eta/S$ approaches zero, proportional to $T^2$,
in the low-temperature limit.

It is worth commenting that the underlying reason for the violation of the
KSS bound
is very different from the standpoint of the boundary field theory versus
the bulk gravity.  For example, for the linear axion models corresponding
to $p=1$, the violation of the bound is attributed in the boundary
field theory to the linear dependence of the axions on the spatial
coordinates, which breaks the translational invariance. On the other hand,
from the viewpoint of the bulk theory,
one may view the translational symmetry as remaining unbroken, at the expense
of the breaking of the axionic shift symmetries of the scalar fields.
(Analogous statements are true also for all $p\ge 1$.)  Furthermore, from
the bulk viewpoint, the full $SO(n-2)=SO(Np)$ rotational symmetry in the
transverse space may be considered to be
preserved for $p=1$.  For $p\ge 3$, the rotational symmetry breaks down to $SO(p)^N$.
In this case, $m(r)^2$ for TT modes within a $p$-dimensional
subspace spanned by a magnetic charge vanishes, while $m(r)^2$ is
non-vanishing if the TT mode encompasses two distinct subspaces.
The solution with a single $p=2$ field strength is more complicated since
the rotational symmetry breaks down to $U(N)$ in this case.  In all cases,
the violation of the KSS bound is related to the non-vanishing of $m(r)^2$;
however, the direct culprit for such a non-vanishing mass in the bulk wave
equation for the TT modes is not entirely clear.  It is also worth commenting
that the magnetically charged solutions break the scaling symmetry
associated with the generalized Smarr relation that is the bulk dual to the KSS bound in a two-derivative gravity theory \cite{llpsma,liu}.

We also derived magnetically-charged Lifshitz black holes with
hyperscaling violation, which we presented in an appendix.

Since the matter fields are spatially dependent, the solutions we considered 
provide simple holographic models for studying momentum relaxation. It
would also be interesting to study the transport properties of these
solutions.

\section*{Acknowledgements}

We are grateful to Sera Cremonini for helpful discussions.
H-S.L.~is supported in part by NSFC grants No. 11305140, 11375153, 11475148 and CSC scholarship No. 201408330017. C.N.P.~is
supported in part by DOE grant DE-FG02-13ER42020. The work of
H.L.~is supported in part by NSFC grants NO. 11475024
and NO. 11235003.

\appendix

\section{Lifshitz black holes with hyperscaling violation}

In this section, we shall construct black hole solutions of the type
(\ref{metricans}), but carrying magnetic charges for form fields of
different ranks.
 The energy-momentum tensors of the field strengths with different ranks
have different fall-off rates, and it is necessary to introduce a
dilaton field that can act as a compensator so that the metric
 in the $(n-2)$ brane directions can still be isotropic.  Thus we
shall consider the Lagrangian
\be
{\cal L} = \sqrt{-g} \Big(R - 2 \Lambda e^{a \phi} -
\fft 12 (\partial \phi)^2 -
 \fft12 \sum_{i=1}^N \fft{1}{p_i !} e^{b_i \phi}\, (F^i_{\sst{(p_i)}})^2
\Big)\,,\label{dillag}
\ee
where $a$ and $b_i$ are dilaton coupling constants. We assume that the
spacetime dimension is
\be
n=2+\sum_{i=1}^N p_i\,.\label{nNp}
\ee
A subset of the theories of the form (\ref{dillag}) comprises those
that are obtained from a circle reduction of the theories (\ref{Npformlag})
that we studied earlier.  For example, if we start from the $n$-dimensional
theory of a single $p$-form coupled to gravity, then the theory in $n-1$
dimensions comprises a $p$-form, a $(p-1)$-form and a dilaton coupled to
gravity. The resulting $(n-1)$-dimensional magnetic brane solution is
also an asymptotically-AdS planar black hole, with $n=N p+2$.  In this
appendix, we shall not give the details of these solutions, since the
dimensional reduction is straightforward.  Instead we consider the more
general Lagrangians that cannot be obtained from such a dimensional reduction.

   We construct static and isotropic black holes with the metric form
(\ref{metricans}), with the ansatz for the metric and
form fields being given by
\bea
d\Sigma^2&=&\sum_{i=1}^N d\Sigma_{i}^2\,,\nn\\
d\Sigma_1^2&=& dx_1^2 + \cdots + dx_{p_1}^2\,,\quad
d\Sigma_2^2 = dy_1^2 + \cdots dy_{p_2}^2\,,\quad \hbox{etc.}\,,\nn\\
\qquad F^i_{(p_i)} &=& \alpha_i\, \Sigma_{i}\,,\nn\\
\Sigma_{1}&=& dx_1\wedge dx_2\wedge\cdots \wedge dx_{p_1}\,,\quad
d\Sigma_{2}= dy_1\wedge dy_2\wedge\cdots \wedge dy_{p_2}\,,\quad
\hbox{etc.}
\eea
The equations of motion for the form fields are automatically satisfied.
The scalar equation is given by
\be
\Box\phi = 2a\Lambda e^{a\phi} +
\sum_{i=1}^N \fft{b_i \alpha_i^2e^{b_i \phi}}{2r^{2p_i}}\,.
\ee
The Einstein equations of motion are
\bea
&&G_{rr} - \phi'^2 -\fft{1}{2f}\Big(-2\Lambda e^{a\phi}
   - \sum_{i=1}^N\fft{\alpha_i^2 e^{b_i\phi}}{2r^{2p_i}}\Big)=0\,,\cr
&&G_{tt} +\ft12 h \Big(-2\Lambda e^{a\phi} -\ft12 \phi'^2
   - \sum_{i=1}^N\fft{\alpha_i^2 e^{b_i\phi}}{2r^{2p_i}}\Big)=0\,,\cr
&&G_{x_i x_i} -\fft{\alpha_i^2 e^{b_i\phi}}{2r^{2p_i-2}} -\ft12 r^2
\Big(-2\Lambda e^{a\phi} -\ft12 \phi'^2-
   \sum_{i=1}^N\fft{\alpha_i^2 e^{b_i\phi}}{2r^{2p_i}}\Big)=0\,,
\eea
where $G_{\mu\nu}=R_{\mu\nu} - \ft12 R g_{\mu\nu}$ is the Einstein tensor.
Since $G_{x_i x_i}$ in the different blocks $d\Sigma_{i}^2$ must the same
for the isotropic solutions, the quantities
\be
\fft{\alpha_i^2 e^{b_i \phi}}{r^{2p_i-2}}\nn
\ee
must be the same for all $i=1,2,\ldots, N$.  For general parameters, this implies that
$\phi=\gamma \log r$ with
\be
\alpha_i=\alpha\,,\qquad \gamma b_i - 2p_i={\rm const.}
\qquad\hbox{for}\qquad i=1,2,\ldots\,N\,.
\ee
The constant $\gamma$ can be determined from the scalar equation,
which implies
\be
\gamma\, b_i - 2p_i = \gamma\,a\,.
\ee
Summing over the all the $i$ indices yields
\be
\gamma=\fft{2(n-2)}{b-N a }\,,\qquad b=\sum_{i=1}^n b_i\,.
\ee
The magnetic charge parameter $\alpha$ is also fully determined,
and is given by
\be
\alpha^2 = \fft{4(-\Lambda)(Na^2 -a b - 2)}{N a b -2N - b^2}\,.\label{alpvalue}
\ee
The metric functions $h$ and $f$ are given by
\be
h=r^{\fft{\gamma^2}{n-2}} f\,,\qquad
f = \sigma \Big(r^{2 + \gamma\, a} -
    \fft{\mu}{r^{n-3 + \fft{\gamma^2}{2(n-2)}}}\Big)\,,\label{hfsol2}
\ee
where
\be
\sigma=\fft{8(n-2)^2(-\Lambda)a}{\gamma \big((n-2)(ab-2) +
\gamma b\big) \big(2(n-1)(n-2) + 2(n-2)a\gamma + \gamma^2\big)}\,.
\ee
Thus the solutions are asymptotic to Lifshitz spacetimes
\cite{lifkachru, liftaylor}, with hyperscaling violation
\cite{Charmousis:2010zz,Iizuka:2011hg,Gouteraux:2011ce,Huijse:2011ef}.
Such black hole solutions in the literature typically involve Maxwell fields
carrying electric
charges, whilst our solutions involve various $p$-forms that carry
magnetic charges.  Note that if $p_i=p$ for all $i$ (with either
a single or else multiple $p$-form field strengths), we must have
$\alpha_i=\alpha$ and $b_i=b$, but $\phi$ does not have to be proportional
to $\log r$.  In particular, $\phi$ can vanish asymptotically as
$r\rightarrow \infty$, giving rise to an AdS planar black hole instead
of a Lifshitz black hole.  Analogous AdS planar black holes also arise
from the Kaluza-Klein reduction of the theory discussed in section 2,
as we commented under (\ref{nNp}).

Note that the solution for the general cases contains only one
integration constant, associated with the mass parameter $\mu$. The magnetic charge parameter $\alpha$ is fixed in terms of the ``cosmological constant'' (\ref{alpvalue}).
When $a=0$, the metric in the Einstein frame is asymptotic to a Lifshitz
spacetime; for non-vanishing $a$, the metric is conformal to a Lifshitz
spacetime, with a hyperscaling violation. To be explicit, we define a new
radial coordinate
\be
\rho=r^{1 + \fft12 \gamma a}\,,
\ee
in terms of which the metric becomes
\bea
ds^2 &=& \rho^{-\fft{2\gamma a}{2+\gamma a}}
d\tilde s^2 = e^{-a\phi} d\tilde s^2\,,\cr
d\tilde s^2 &=& -\rho^{\fft{2(\gamma^2 + (n-2)\gamma a)}{(n-2)(2+\gamma a)}}
\tilde f dt^2 + \fft{d\rho^2}{\tilde f} + \rho^2 d\Sigma^2\,,\cr
\tilde f &=& \sigma (1 + \ft12 \gamma a)^2
\Big[\rho^2 - \mu\, \rho^{-\fft{2(n-2)(n-3) +
\gamma^2}{(n-2)(2 + \gamma a)}}\Big]\,.
\eea

We find that the transverse and traceless modes also satisfy the wave equation
\be
(\Box - m(r)^2)\Psi(r,t)=0\,,
\ee
but now with $m(r)^2$ given by
\be
m(r)^2 = \nu \alpha^2\, r^{\gamma a}\,.
\ee
It is clear that when $\gamma a\ge 0$, the mass is non-vanishing
as $r\rightarrow \infty$. It follows that the zero-frequency mode
$\chi$ cannot approach a constant as $r\rightarrow \infty$. In order
for $\chi$ to approach 1 asymptotically, as in the case we discussed in
the main text, it is necessary to have $m(r)$ vanish on the boundary with
a sufficiently rapid fall off.  However, the solution (\ref{hfsol2})
implies that we must have $\gamma a>-2$.  This makes it impossible
that $\chi$ can approach 1 as $r\rightarrow \infty$.  The implications for
the dual boundary theory in this case will be addressed in
a future publication.



\begin{thebibliography}{99}

\bibitem{dhokkrau} E. D'Hoker and P. Kraus,
{\it Magnetic brane solutions in AdS},
  JHEP {\bf 0910}, 088 (2009), arXiv:0908.3875 [hep-th].

\bibitem{andrade}
  T. Andrade and B. Withers,
  {\it A simple holographic model of momentum relaxation,}
  JHEP {\bf 1405}, 101 (2014),  arXiv:1311.5157 [hep-th].

\bibitem{harramsan} S.A. Hartnoll, D.M. Ramirez and J.E. Santos,
{\it Entropy production, viscosity bounds and bumpy black holes},
  arXiv:1601.02757 [hep-th].

\bibitem{adscft1} J.M. Maldacena,
{\it The large N limit of superconformal field theories and supergravity},
Adv.\ Theor.\ Math.\ Phys.\  {\bf 2}, 231 (1998),
hep-th/9711200.

\bibitem{adscft2} S.S. Gubser, I.R. Klebanov and A.M. Polyakov,
{\it Gauge theory correlators from non-critical string theory},
Phys.\ Lett.\  {\bf B428}, 105 (1998),
hep-th/9802109.

\bibitem{adscft3} E. Witten,
{\it Anti-de Sitter space and holography},
Adv. Theor. Math. Phys. {\bf 2}, 253 (1998), hep-th/9802150.

\bibitem{adscft4}
  O. Aharony, S.S. Gubser, J.M. Maldacena, H. Ooguri and Y. Oz,
{\it Large $N$ field theories, string theory and gravity,}
  Phys.\ Rept.\  {\bf 323}, 183 (2000)
  [hep-th/9905111].

\bibitem{Policastro:2001yc}
  G. Policastro, D.T. Son and A.O. Starinets,
{\it The shear viscosity of strongly coupled ${\cal N}=4$
supersymmetric Yang-Mills plasma,}
  Phys.\ Rev.\ Lett.\  {\bf 87}, 081601 (2001),
 hep-th/0104066.

\bibitem{sonsta} D.T. Son and A.O. Starinets,
{\it Minkowski space correlators in AdS/CFT correspondence:
Recipe and applications},
JHEP {\bf 0209}, 042 (2002), hep-th/0205051.

\bibitem{KSS} P. Kovtun, D.T. Son and A.O. Starinets,
{\it Holography and hydrodynamics: Diffusion on stretched horizons},
JHEP {\bf 0310}, 064 (2003), hep-th/0309213.

\bibitem{KSS0} P. Kovtun, D.T. Son and A.O. Starinets,
{\it Viscosity in strongly interacting quantum field theories from black
hole physics},  Phys.\ Rev.\ Lett.\  {\bf 94}, 111601 (2005),
hep-th/0405231.

\bibitem{llpsma} H.S. Liu, H. L\"u and C.N. Pope,
{\it Generalized Smarr formula and the viscosity bound for
Einstein-Maxwell-dilaton black holes},
  Phys.\ Rev.\ D {\bf 92}, no. 6, 064014 (2015),
arXiv:1507.02294 [hep-th].

\bibitem{liu}
  H.S. Liu,
  {\it Global scaling symmetry, Noether charge and
universality of shear viscosity,}
  arXiv:1601.07875 [hep-th].

\bibitem{kats}
  Y. Kats and P. Petrov,
 {\it Effect of curvature squared corrections in AdS on the viscosity of
the dual gauge theory,}
  JHEP {\bf 0901}, 044 (2009),  arXiv:0712.0743 [hep-th].

\bibitem{brlimyshya} M. Brigante, H. Liu, R.C. Myers, S. Shenker and S. Yaida,
{\it Viscosity bound violation in higher derivative gravity},
Phys.\ Rev.\ D {\bf 77}, 126006 (2008), arXiv:0712.0805 [hep-th].

\bibitem{caniohsu} R.G. Cai, Z.Y. Nie, N. Ohta and Y.W. Sun,
{\it Shear viscosity from Gauss-Bonnet gravity with a dilaton coupling},
  Phys.\ Rev.\ D {\bf 79}, 066004 (2009),
arXiv:0901.1421 [hep-th].

\bibitem{cremonini} S. Cremonini,
{\it The shear viscosity to entropy ratio: a status report},
  Mod.\ Phys.\ Lett.\ B {\bf 25}, 1867 (2011),
arXiv:1108.0677 [hep-th].

\bibitem{Natsuume:2010ky}
  M. Natsuume and M. Ohta,
{\it The shear viscosity of holographic superfluids,}
  Prog.\ Theor.\ Phys.\  {\bf 124}, 931 (2010),
 arXiv:1008.4142 [hep-th].

\bibitem{Rebhan}
  A. Rebhan and D. Steineder,
  {\it Violation of the holographic viscosity bound in a strongly coupled
anisotropic plasma,}
  Phys.\ Rev.\ Lett.\  {\bf 108}, 021601 (2012), arXiv:1110.6825 [hep-th].

\bibitem{Erdmenger:2010xm}
  J. Erdmenger, P. Kerner and H. Zeller,
{\it Non-universal shear viscosity from Einstein gravity,}
  Phys.\ Lett.\ B {\bf 699}, 301 (2011),
  arXiv:1011.5912 [hep-th].

\bibitem{mamo}
  K.A. Mamo,
  {\it Holographic RG flow of the shear viscosity to entropy density ratio
in strongly coupled anisotropic plasma,}
  JHEP {\bf 1210}, 070 (2012),  arXiv:1205.1797 [hep-th].

\bibitem{Critelli:2014kra}
  R. Critelli, S.I. Finazzo, M. Zaniboni and J. Noronha,
{\it Anisotropic shear viscosity of a strongly coupled non-Abelian plasma from magnetic branes,}
  Phys.\ Rev.\ D {\bf 90}, no. 6, 066006 (2014)
  doi:10.1103/PhysRevD.90.066006
  [arXiv:1406.6019 [hep-th]].


\bibitem{Bhattacharyya:2014wfa} A. Bhattacharyya and D. Roychowdhury,
{\it Viscosity bound for anisotropic superfluids in higher derivative gravity,}
  JHEP {\bf 1503}, 063 (2015),
arXiv:1410.3222 [hep-th].

\bibitem{Ge:2014aza}
X.H. Ge, Y. Ling, C. Niu and S.J. Sin,
{\it Holographic transports and stability in anisotropic linear axion model,}
  arXiv:1412.8346 [hep-th].

\bibitem{fllp1}
  X.H. Feng, H.S. Liu, H. L\"u and C.N. Pope,
 {\it Black hole entropy and viscosity bound in Horndeski gravity,}
  JHEP {\bf 1511}, 176 (2015),   arXiv:1509.07142 [hep-th].

\bibitem{fllp2}
  X.H. Feng, H.S. Liu, H. L\"u and C.N. Pope,
 {\it Thermodynamics of charged black holes in Einstein-Horndeski-Maxwell
theory,}
  Phys.\ Rev.\ D {\bf 93}, 044030 (2016), arXiv:1512.02659 [hep-th].


\bibitem{alberte}
  L. Alberte, M. Baggioli and O. Pujolas,
 {\it Viscosity bound violation in holographic solids and the
viscoelastic response,}
  arXiv:1601.03384 [hep-th].

\bibitem{burikham}
  P. Burikham and N. Poovuttikul,
  {\it Shear viscosity in holography and effective theory of transport
without translational symmetry,}
  arXiv:1601.04624 [hep-th].

\bibitem{Bardoux:2012aw}
  Y. Bardoux, M.M.~Caldarelli and C. Charmousis,
{\it Shaping black holes with free fields,}
  JHEP {\bf 1205}, 054 (2012)
  [arXiv:1202.4458 [hep-th]].

\bibitem{taywoo} M. Taylor and W. Woodhead,
{\it Inhomogeneity simplified},
  Eur.\ Phys.\ J.\ C {\bf 74}, no. 12, 3176 (2014),
arXiv:1406.4870 [hep-th].

\bibitem{Ortaggio:2007hs}
  M.~Ortaggio, J.~Podolsky and M.~Zofka,
{\it Robinson-Trautman spacetimes with an electromagnetic field in higher dimensions,}
  Class.\ Quant.\ Grav.\  {\bf 25}, 025006 (2008)
  doi:10.1088/0264-9381/25/2/025006
  [arXiv:0708.4299 [gr-qc]].

\bibitem{Ortaggio:2014gma}
  M. Ortaggio, J. Podolsky and M. Zofka,
{\it Static and radiating p-form black holes in the higher dimensional Robinson-Trautman class,}
  JHEP {\bf 1502}, 045 (2015)
  [arXiv:1411.1943 [gr-qc]].

\bibitem{Donos:2011qt}
  A. Donos, J.P. Gauntlett and C. Pantelidou,
{\it Spatially modulated instabilities of magnetic black branes,}
  JHEP {\bf 1201}, 061 (2012)
  [arXiv:1109.0471 [hep-th]].

\bibitem{Cvetic:1999xp}
  M. Cveti\v c {\it et al.},
{\it Embedding AdS black holes in ten-dimensions and eleven-dimensions,}
  Nucl.\ Phys.\ B {\bf 558}, 96 (1999)
  [hep-th/9903214].

\bibitem{Davison:2014lua}
  R.A. Davison and B. Gouteraux,
{\it Momentum dissipation and effective theories of coherent and
incoherent transport,}
  JHEP {\bf 1501}, 039 (2015), arXiv:1411.1062 [hep-th].

\bibitem{buccre} A. Buchel and S. Cremonini,
{\it Viscosity bound and causality in superfluid plasma},
  JHEP {\bf 1010}, 026 (2010),
arXiv:1007.2963 [hep-th].

\bibitem{lifkachru}
  S. Kachru, X. Liu and M. Mulligan,
  {\it Gravity duals of Lifshitz-like fixed points,}
  Phys.\ Rev.\ D {\bf 78}, 106005 (2008),  arXiv:0808.1725 [hep-th].

\bibitem{liftaylor}
  M. Taylor,
  {\it Non-relativistic holography,}   arXiv:0812.0530 [hep-th].

\bibitem{Charmousis:2010zz}
  C. Charmousis, B. Gouteraux, B. S. Kim, E. Kiritsis and R. Meyer,
{\it Effective holographic theories for low-temperature condensed matter
systems},
  JHEP {\bf 1011}, 151 (2010),
 arXiv:1005.4690 [hep-th].

\bibitem{Iizuka:2011hg}
  N. Iizuka, N. Kundu, P. Narayan and S.P. Trivedi,
{\it Holographic Fermi and non-Fermi liquids with transitions in
dilaton gravity},
  JHEP {\bf 1201}, 094 (2012),
  arXiv:1105.1162 [hep-th].

\bibitem{Gouteraux:2011ce}
  B. Gouteraux and E. Kiritsis,
{\it Generalized holographic quantum criticality at finite density},
  JHEP {\bf 1112}, 036 (2011), arXiv:1107.2116 [hep-th].

\bibitem{Huijse:2011ef}
  L. Huijse, S. Sachdev and B. Swingle,
{\it Hidden Fermi surfaces in compressible states of gauge-gravity duality},
  Phys.\ Rev.\ B {\bf 85}, 035121 (2012), arXiv:1112.0573 [cond-mat.str-el].

\end{thebibliography}
\end{document}